\documentclass[12pt,showpacs]{revtex4}

\usepackage{setspace}
\usepackage{times}
\usepackage{graphicx}

\def\gtorder{\mathrel{\raise.3ex\hbox{$>$}\mkern-14mu
 \lower0.6ex\hbox{$\sim$}}}
\def\ltorder{\mathrel{\raise.3ex\hbox{$<$}\mkern-14mu
 \lower0.6ex\hbox{$\sim$}}}

\def\ge{G_E}
\def\gm{G_M}

\def\mugegm{\mu_p G_E / G_M}

\begin{document}

\title{An examination of proton charge radius extractions from e-p scattering data}



\author{John Arrington}
\affiliation{Physics Division, Argonne National Laboratory, Argonne, IL 60439}

\begin{abstract}

A detailed examination of issues associated with proton radius extractions
from elastic electron-proton scattering experiments is presented. Sources of
systematic uncertainty and model dependence in the extractions are discussed,
with an emphasis on how these may impact the proton charge and magnetic radii.
A comparison of recent Mainz data to previous world data is presented,
highlighting the difference in treatment of systematic uncertainties as well
as tension between different data sets. We find several issues that suggest
that larger uncertainties than previously quoted may be appropriate, but do
not find any corrections which would resolve the proton radius puzzle.

\end{abstract}

\maketitle


\section{Introduction}

\begin{figure}[tbp]
\includegraphics[angle=0,width=11.0cm]{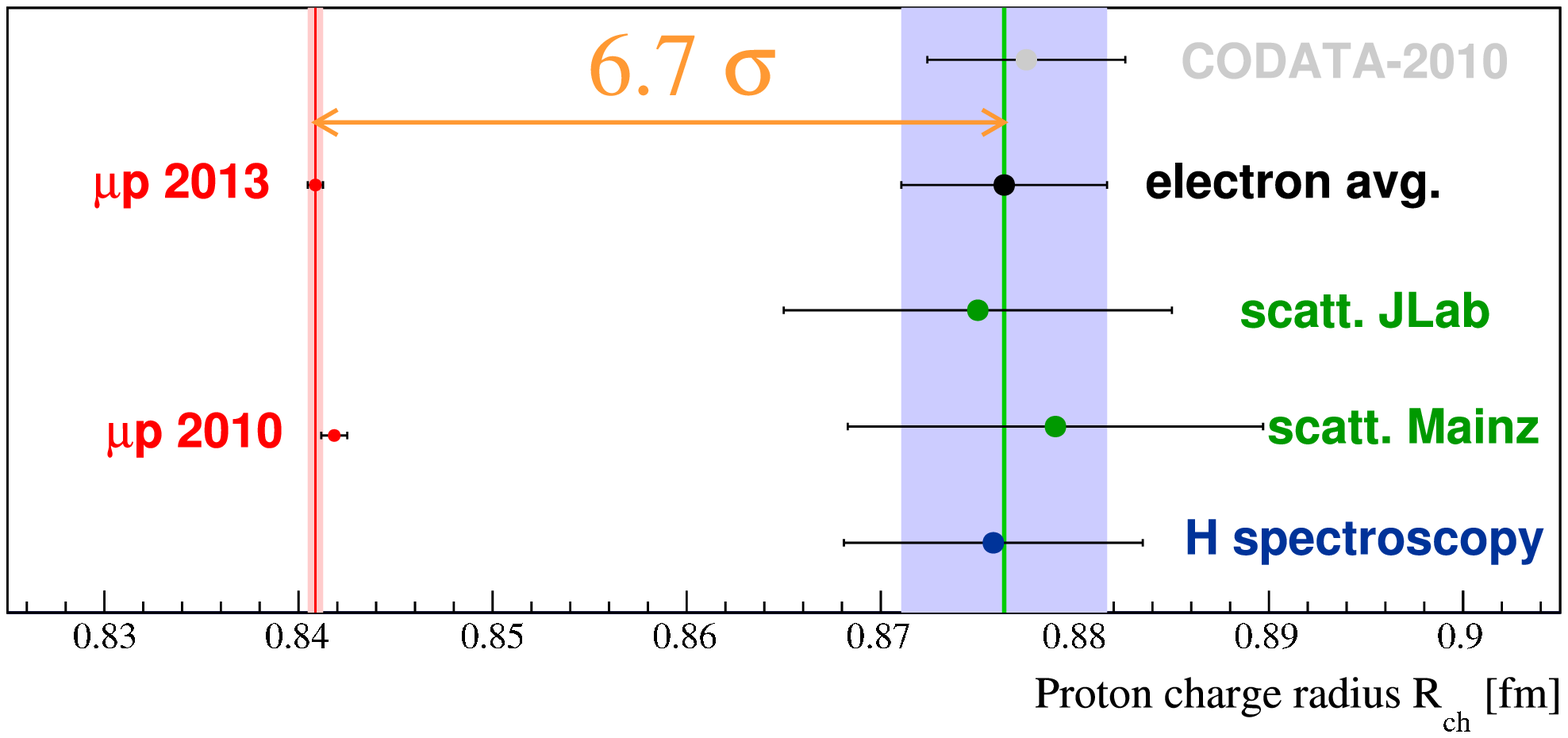}
\caption{Extractions of the proton charge radius from muonic hydrogen
measurements~\cite{pohl10, antognini13}, hydrogen spectroscopy~\cite{mohr12},
electron scattering measurements at Mainz~\cite{bernauer13, uncertainties},
and a global analysis of earlier world data~\cite{zhan11}. The direct average
shown is compared to the CODATA-2010 evaluation~\cite{mohr12}. Figure courtesy
of Randolf Pohl.}
\label{fig:radii}
\end{figure}

Five years after the initial extraction of the proton radius from muonic
hydrogen~\cite{pohl10}, the ``proton radius puzzle'' persists. 
Measurements based on muonic hydrogen transitions~\cite{antognini13} and those
based on electron transitions~\cite{mohr12} or electron scattering
measurements~\cite{zhan11, sick12, bernauer13} disagree at the 7$\sigma$
level, with muonic hydrogen results yielding a radius near 0.84~fm and 
electron-based measurements yielding $r_E \approx 0.88$fm, as summarized in
Fig.~\ref{fig:radii}. In light of this, a
careful examination of the details of these extractions is clearly warranted.
Here, we discuss several issues relevant to determining the proton radius
from electron scattering data.

In examining extractions from electron scattering data, we examine the
Mainz data~\cite{bernauer13} and global analyses~\cite{zhan11, sick12} of
world data (excluding Mainz) separately. This is done because the Mainz data
presents the uncertainties in the data in a significantly different way from
most other experiments, making it difficult to perform a meaningful combined
analysis. It is also beneficial to perform independent analyses to examine
consistency between the Mainz data and other measurements at the cross
sections level, which can be overlooked in a combined analysis. We also discuss
some preliminary results from a detailed examination of both Mainz and world
data~\cite{lee15}. There are several issues that suggest that larger
uncertainties than quoted in previous works are warranted. While none of these
appear likely to resolve to the discrepancy with muonic hydrogen
measurements, some issues remain which deserve more detailed examination.

\section{General issues in the extraction of the radii}

One obtains the charge and magnetic form factors, $G_E(Q^2)$ and $G_M(Q^2)$,
from unpolarized cross section measurements by performing a Rosenbluth
separation~\cite{rosenbluth50} which uses the angle-dependence at fixed $Q^2$
to separate the charge and magnetic contributions. The cross section at fixed
$Q^2$ is proportional to the 'reduced' cross section $\sigma_R = \tau G_M^2 +
\varepsilon G_E^2$, where $\tau = Q^2/(4M_p^2)$ and
$\varepsilon^{-1}=[1+2(1+\tau)\tan^2(\theta/2)]$. At low $Q^2$, the magnetic
contribution is strongly suppressed except for very small $\varepsilon$ values,
corresponding to large scattering angle. Because of the difficulties in making
very large angle scattering measurements at low $Q^2$, a significant
extrapolation to $\varepsilon=0$ is required and even sub-percent
uncertainties on the cross sections can yield significant uncertainties on
small contribution from $G_M(Q^2)$.

Because one often combines data from many experiments, each of which
has an uncertainty in its normalization uncertainty, the normalizations 
factors of the limited number of large-angle data sets have a great impact on
the extraction of $G_M$. If these normalization factors are allowed to vary in
the fit, which is the most common approach, then a small shift in
normalization between large and small angle data sets can yield a significant
shift of strength between $G_E$ and $G_M$ over a range in $Q^2$ values.
Polarization observables are sensitive to the ratio
$G_E/G_M$~\cite{perdrisat07, arrington07a} and can thus provide not only
direct information on the form factors, but also improve the determination of
the relative normalization of different measurements. Of particular interest
are data sets at low $Q^2$ values~\cite{ron07, blast, zhan11, ron11} which
provide improved extractions of $\gm$ and additional constraints on the
experimental normalizations.

Extraction of the charge radius from electron scattering requires
parameterizing the cross sections to obtain the slope of the form factor at
$Q^2=0$. Many naive extractions use fit functions which do not provide
sufficient flexibility to accurately describe the low-$Q^2$ data, and often
do not attempt to estimate the uncertainty associated with the choice of 
functional form used to fit the data. For example, several early extractions
were based on linear fits to low $Q^2$ data. Such fits will always give an
underestimate of the radius, based on the observed positive curvature of the
form factors at low $Q^2$. One would have to have extremely precise data at
very low $Q^2$ for a linear fit to be sufficient~\cite{sick03, kraus14}.

Simply increasing the number of parameters in a simple Taylor expansion or
similar fit function provides greater flexibility. However, it also leads to
increased uncertainty in the extracted radius as correlations between
different parameters in the expansion allow the impact of variations in one term of the
expansion to be balanced by changes in other terms. This yields a rapid increase in the
extracted radius uncertainty as the number of parameters is increased. This
may lead to a situation where there is no region in which there are enough
parameters to accurately reproduce the data while still yielding an
uncertainty small enough to provide a useful radius extraction~\cite{lee15}.
Such analyses must find a balance between fit flexibility and radius
uncertainty and, ideally, attempt to estimate the error made when truncating
the fit function. All of the extractions that we will review in
detail~\cite{zhan11, sick12, bernauer13} examine the model dependence
associated with the functional form used to parameterize the data and include
at least some estimate of the associated uncertainty.

\begin{figure}[tbp]
\includegraphics[angle=0,width=7.8cm]{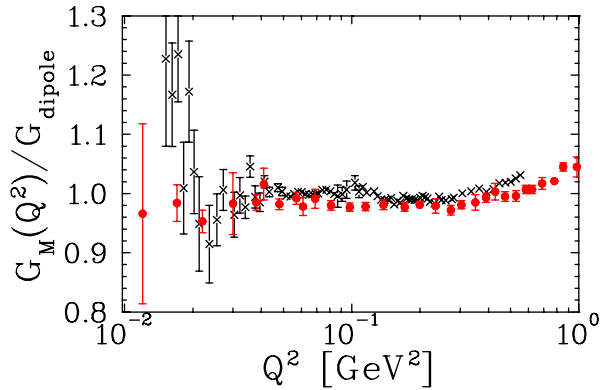}
\caption{Extractions of $\gm$ and their uncertainties from direct Rosenbluth
separations for the Mainz data~\cite{bernauer13} (crosses) and from a global
(pre-Mainz) analysis~\cite{arrington07c} (circles).}
\label{fig:gmp}
\end{figure}

The factor $\tau$ suppresses the magnetic contribution as $Q^2 \to 0$, causing
the uncertainties on $G_M(Q^2)$ to increase rapidly as seen in
Fig.~\ref{fig:gmp}. Because the radius extraction is sensitive to the
low-$Q^2$ behavior of $\gm$ and the most precise data are at higher $Q^2$
values, it is particularly difficult to reliably extract the magnetic radius.
In the analyses of world data~\cite{zhan11, sick12}, the fits exclude
high-$Q^2$ data to prevent these data from influencing the extraction of the
slope. For the analysis of the Mainz data~\cite{bernauer13}, the data set
extends to $Q^2 \approx 1$~GeV$^2$, but the bulk of the data below
$Q^2=0.5$~GeV$^2$ and a more flexible fit function is used to provide greater
flexibility to fit cross section measurements at both low and high $Q^2$.

Radiative corrections are another area requiring special attention. The
largest contributions to the radiative corrections can be calculated in a
model-independent way, although there are small variations between different
prescriptions~\cite{mo69, maximon00, vanderhaeghen00, ent01}. Other terms,
in particular the two-photon exchange (TPE) contributions~\cite{carlson07,
arrington11b}, are model dependent as it is necessary to account for the
possible hadronic states in between the two exchanged photons. The world
data analyses~\cite{zhan11, sick12} include two-photon exchange corrections
based on a calculation in a hadronic basis including only intermediate proton
state~\cite{blunden05a}, although estimates of excited
states~\cite{kondratyuk05, kondratyuk07} suggest that their contribution is
very small at the relevant $Q^2$ values. The radiative correction
uncertainties quoted by the experiments used in these global analyses were
typically 1-1.5\%, and are assumed to be sufficient after applying the
calculate TPE corrections. We note that the data of Simon, et
al.~\cite{simon80,simon81} did not include any uncertainty for radiative
corrections and thus tend to have an artificially enhanced impact on
extractions of the form factors and radius. In the analysis of
Ref.~\cite{zhan11}, and additional radiative correction uncertainty was
applied to the Simon data.

The primary result from the Mainz experiment~\cite{bernauer13} applies TPE
corrections derived for a point target~\cite{mckinley48} (the ``Feshbach'' 
correction). This correction is exact for $Q^2=0$ but has no $Q^2$ dependence.
Because the radius is the $Q^2$ slope of the form factors at $Q^2=0$, it seems
unlikely that a $Q^2$-independent correction will be sufficient. The
model-dependent TPE calculations mentioned above agree with the Feshbach
correction at $Q^2=0$, but as $Q^2$ increases they tend to decrease, 
going to zero before changing sign and growing in magnitude above
$Q^2 \approx 0.3$~GeV$^2$. There are several TPE calculations meant to be
appropriate at low $Q^2$~\cite{blunden05a, borisyuk06, borisyuk08,
kondratyuk05, borisyuk07, borisyuk12, borisyuk13}, and they are all in good
agreement at low $Q^2$ as shown in Ref.~\cite{arrington13}. Very recently,
this change of sign relative to the $Q^2=0$ limit was confirmed by comparisons
of electron-proton and positron-proton scattering for $Q^2 \approx 1$ and
1.5~GeV$^2$~\cite{adikaram15, rachek15}. This supports the idea that the
$Q^2=0$ calculation is not appropriate and a more complete TPE correction is
required.

The question of TPE corrections in the Mainz data was first examined in
Refs.~\cite{arrington11c, bernauer11}. Ref.~\cite{bernauer11} shows a direct
comparison of the extracted value of $\mugegm$ with and without TPE
corrections from Ref.~\cite{borisyuk07}, which are expected to be valid up to
$Q^2 \approx 0.1$~GeV$^2$. As noted in~\cite{bernauer11}, the correction on
$\mugegm$ is relatively small, below 1\%. However, this correction is larger
than the linear sum of the statistical, systematic, and model uncertainties.
It is, therefore, a critical correction for an extraction aimed at such high
precision, and clearly necessary for a precise extraction of the charge and
magnetic radii. Ref.~\cite{bernauer13} does not include any uncertainty
associated with TPE corrections, but does include an extraction of the radius
after applying hadronic corrections with the proton intermediate
state~\cite{blunden05a}. The change in the charge radius is 0.004~fm, roughly
1/3 of the total uncertainty~\cite{uncertainties}, while the magnetic radius
changes by 0.022~fm, more than the total quoted uncertainty.

\section{Examination of the Mainz analysis}

As noted earlier, the extraction of the uncertainties as well as the
breakdown of different types of uncertainties in the recent Mainz data set
is significantly different from other experiments. We describe the approach
used in Ref.~\cite{bernauer13}, and then discuss potential implications on the
uncertainties of the extracted radii in the Mainz analysis, as well as 
independent fits to the Mainz cross section data.

\subsection{Uncorrelated systematic uncertainty}

The uncorrelated systematic uncertainties were determined by performing a fit
to the full data set using only the pure counting statistics for uncertainties.
The difference between the data and fit for each subset (each independent
energy-spectrometer combination) was examined, and a scaling factor was 
determined for each data set which, when applied as a scale factor enhancement
to the uncertainties from the counting statistics on every data point, yielded
a scatter that was approximately consistent with the enhanced statistical
uncertainty. The goal is to provide a reduced $\chi^2$ value closer to unity,
with $\chi^2_\nu \approx 1.14$ for the final Mainz fit to the cross sections
with the scaling factors applied. This procedure yields the minimal
uncertainty necessary to account for the non-statistical scatter of the data,
but is insensitive to any sources of error which may be correlated with the
kinematics or operating conditions of the experiment, e.g. beam energy or
spectrometer angle offsets, approximations in the radiative correction
procedures, or subtraction of target cell wall contributions. In fact, because
the final reduced chi-squared is still above one, the final uncorrelated
systematic uncertainty is somewhat below the minimum necessary to account for
the observed scatter.

This rescaling procedure is relatively unusual; nearly all other experiment
made direct estimates of uncertainties or upper limits for various
sources of uncertainty which may be treated as uncorrelated in the fit. This
uncorrelated systematic is determined and added in quadrature to the
statistical uncertainty. If we convert the Mainz scaling factors to 
independent systematic uncertainties using this standard approach, they
correspond to uncertainties that average 0.25\%, but vary from 0.02\% to 2\%
with the smallest systematic uncertainties generally being applied to the data
with smallest statistical uncertainties.

\subsection{Correlated systematic uncertainty}

Most experiments provide relatively small data sets, typically tens of cross
section measurements covering a range of $Q^2$ and $\varepsilon$ values. For
such data sets, correlated errors, e.g. associated with kinematic-dependent
corrections, can be well represented by applying an additional uncorrelated
uncertainty to each point. A modest 0.5\% contribution to the uncertainty on
each cross section provides flexibility to cover an arbitrary correlated
uncertainty at the few tenths of a percent level. For the Mainz measurement
there are 1422 cross sections, 10--50 times more than most experiments, so any
limited kinematic region will have many more data points, reducing the impact
of the uncorrelated uncertainty by the square root of the number of
points. Thus, trying to represent small correlated effects as uncorrelated
uncertainties would require much larger contributions.

This effect is made worse by the fact that of the 1422 data points, there are
only 638 independent kinematic settings. In several cases, multiple repeated
measurements were taken at the same kinematic setting, one after the other.
For the given procedure - inflating the counting statistics by a scaling
factor intended to yield a reasonable chi-squared for each data set - it 
doesn't matter that there are multiple repeated measurements in the data set.
However, using the more conventional approach of applying a fixed
systematic uncertainty to each point, a set of $N$ repeated measurements would
artificially reduce the impact of the systematic uncertainty by a factor of
$\sqrt{N}$ for this kinematic setting. If the data are rebinned into their 638
independent points and a systematic applied that yields a reduced chi-squared
value near unity, the uncertainties tend to increase more where the scaled
uncertainties were very small, i.e. the very high statistics points or the
kinematics with small scaling factors. This ends up increasing the low $Q^2$
data uncertainties more and yields a larger uncertainty on the extracted
radius~\cite{lee15}.

\begin{figure}[tbp]
\includegraphics[angle=0,width=7.8cm]{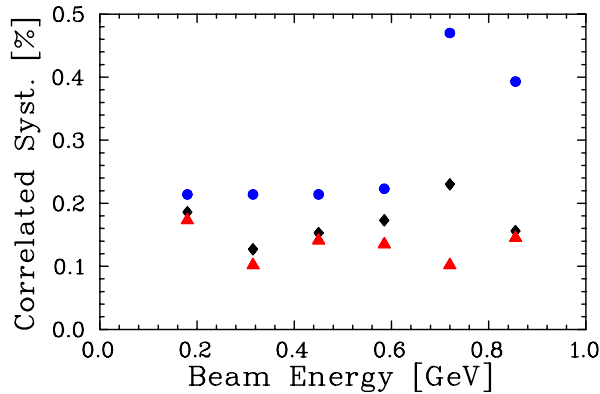}
\caption{Size of the correlated systematic parameter $a$ for Spectrometer
A (diamond), B (triangle), and C (circle) for each beam energy setting of
Ref.~\cite{bernauer13}.}
\label{fig:corr_sys}
\end{figure}

Because of the limitations of including only uncorrelated systematic
uncertainties on such a large data set, the A1 collaboration treated
correlated systematic uncertainties independently. They separated
the full data set into 18 subsets, each corresponding to a single spectrometer
and fixed beam energy. They then simultaneously applied a correction factor,
proportional to the scattering angle, to all points with each of these 18
subgroups and refit the data. The correction varied from 0\% at the smallest
angle to a setting-dependent maximum value at the largest angle for each
setting. The size of this maximum correction, the parameter ``$a$'' from
Ref.~\cite{bernauer13}, is shown for all 18 spectrometer/energy combinations
in Fig.~\ref{fig:corr_sys}, with most settings having a value between 0.1\%
and 0.25\%. In addition, that there is also a separate correlated systematic
which accounts for variation of the cross section with the elastic tail cut,
which is evaluated separately and then combined with correlated systematic
mentioned above. As this is the smaller contribution, we focus here on the
correction that is taken to be linear in $\theta$.

Note the in the supplemental material of Ref.~\cite{bernauer13}, the systematic
correction does not go to zero for the smallest scattering angle of each
subset, but goes to zero at $\theta=0$. If the normalization factors of the
data subgroups are allowed to vary, the two procedures are equivalent. However,
if the normalizations are not allowed to vary, applying the corrections factors
as published will yield a much larger correlated error.

\subsection{Normalization uncertainty}

The combination of the uncorrelated and correlated uncertainties applied
to the Mainz data are extremely small compared to other measurements and, by
themselves, represent an incomplete estimate of the experimental
uncertainty. The data are broken up into 34 subgroups, each of which
has an independent, unconstrained normalization factor. Thus, the
uncorrelated and correlated systematic uncertainties described above need only
account for the \textit{variation} of any corrections over the kinematics of
the individual subgroups which consist of between 18 and 68 independent
kinematics each (treating multiple runs at identical kinematics as single
points), as the overall normalization factor will account for any average
correction. Note, however, that there are 34 different normalization subgroups
while the correlated systematic uncertainties are applied over the 18 
independent beam energy-spectrometer combinations. It would seem more 
consistent for apply the correlated systematic over each of the 34 
normalization subgroups, and the potential impact of this, as well as the
choice of functional form for the correlated systematic, will be discussed
in the following section.

\subsection{Treatment of the uncertainties}

Because the uncertainties of the measurement are separated into
uncorrelated, correlated, and normalization uncertainties, a proper evaluation
of the radius uncertainties must account for all of these. Fits which take the
Mainz cross section data as quoted and do not allow the normalization factors
to vary, e.g. as in Ref.~\cite{lorenz14}, will yield artificially small
uncertainties in the radius extractions, as discussed in Ref.~\cite{lee15}.

In addition, questions have been raised about missing contributions to the
uncertainties. No uncertainties associated with TPE corrections are included,
and there are additional uncertainties associated with approximations made in
radiative correction procedures. These are neglected in~\cite{bernauer13},
based on the assumption that these will be contained in the small correlated
systematic uncertainties applied for other effects, but no argument is made to
support these uncertainties being negligible compared to the typical 0.1-0.2\%
correlated systematic (Fig.~\ref{fig:corr_sys}). They also quote uncertainties
on the knowledge of the beam energy and spectrometer scattering angles, but
do not account for these in the cross section uncertainties. While the impact
of these kinematic uncertainty on the cross sections is generally very small,
the corrections for both energy and angle can be as large as 0.2\%, mainly
at low energy and small scattering angle, and are strongly kinematic
dependent. Thus, it is not clear that they should be neglected in comparison
to statistical uncertainties at the 0.2\% level and correlated systematics
which are as low as 0.1\% for some settings.

It has also been noted~\cite{sick12} that the target cell wall contributions,
subtracted from the data based on a calculated spectrum, don't match the
observed contribution. Based on the visible difference between the data and
simulated spectrum in the region between the nuclear elastic and e-p elastic
peaks (Fig. 8b of Ref.~\cite{bernauer13}), it was estimated that this
subtraction underestimates the cell contributions by 1.2\%~\cite{sick12}. This
is a large effect, both compared to the quoted uncertainties of the
measurement and to the size of the subtraction (below 4\% for most
settings~\cite{bernauer11}). While this is the largest contribution to the
correlated systematic uncertainty, it is still taken to be roughly
0.1-0.2\% for nearly all settings. Any overall normalization error caused by
an underestimate of the background subtraction will be removed in the fit, but
with the subtraction varying from a few percent to nearly 10\% for
spectrometer B at forward angles~\cite{bernauer11}, it's not clear that the
kinematic variation over each of the data subsets can be constrained to be
below the 0.1-0.2\% level.

In addition to noting that the correlated systematic uncertainties applied
are very small and may neglect important contributions, we also note that the
impact of these uncertainties on the extracted form factors and radii are
evaluated within a single model; a shift of the cross sections which is linear
in $\theta$ over each of the 18 spectrometer-beam energy combinations.  Taking
the correction to scale with quantities other than $\theta$ can yield larger
or smaller corrections, with an increase 50\% or more in some models where the
correction is not linear in angle~\cite{lee15}. Note that the procedure always
yields a fixed correction between the smallest and largest angle settings of a
given data subset, so it is only the form of the variation over this subset
that is changed in these tests. The impact of the correlated systematic
uncertainty is also significantly increased if it is taken as a correction to
each data set with independent normalization, which seems a more
consistent approach given the breakdown of uncertainties into uncorrelated,
correlated, and normalization. Given the issues with the size of the
correlated systematics noted above and the model dependence of converting
these systematic effects into an uncertainty on the extracted radius, a
conservative approach would appear to yield significantly larger radius
uncertainties associated with the correlated systematics.

Note that the detailed comparisons in Ref.~\cite{lee15} are performed for the
rebinned version of the Mainz data. When examining the original analysis
procedure~\cite{bernauer13}, the correlated systematic uncertainty on the
extracted radii is more sensitive to both the functional form chosen to
represent the correlated corrections and to the question of whether the
correction is applied to each normalization subgroup or each beam-spectrometer
combination.

\section{Fitting with bounded $z$ expansion}

As noted earlier, it is important that the fit function be flexible enough to
adequately reproduce the data without being so flexible that it does not
provide a meaningful constraint on the extracted radii. This is
especially critical for the magnetic radius, where the precise
high-$Q^2$ data can influence the fit more than the low-$Q^2$ data which
are directly sensitive to the radius. Typically, a fit function is selected and
the number of parameters is chosen to be as small as possible while still
providing a reduced chi-squared that is close to the minimum value obtained
for many parameters. In some cases, fit functions designed to help minimize
the impact of the high-$Q^2$ data on the low-$Q^2$ fit are chosen, e.g. spline
functions~\cite{bernauer13} or continued fraction fits~\cite{sick03}. However,
it is still difficult to determine how many parameters are sufficient for a
reliable fit. Figure 9.21 of Ref.~\cite{bernauerphd} shows the extracted
charge and magnetic radii vs. the number of parameters for a variety of
different functional forms. For the charge radius, the extracted radii are
relatively consistent for fits with 10-15 parameters, although there is a
significant spread ($\approx 0.02$~fm) between the values using different fit
functions. For more than 16 parameters, the spline fits yield much smaller
radii, but this is presumably in the region where the uncertainties become
very large and so the shift of the central value may not be outside of the fit
uncertainty. For the magnetic radius, the situation is noticeably worse. There
is a narrow window, 6-8 parameters, where the radii are relatively stable and
then by 10 parameters the different functional forms yield radius values
differing by nearly 0.2~fm. As noted in Ref.~\cite{sick14}, some of this
erratic behavior is associated with fits that have unphysical behavior, e.g.
poles in the form factor and oscillatory behavior of the proton charge
density, $\rho(r)$ at very large values of $r$.

One approach to this problem was presented in Refs.~\cite{sick12, sick14},
where a parameterization was chosen that constrains the large-$r$ behavior of
the form factors. By including physical constraints on the behavior of the
form factors, one can avoid the possibility that insufficiently flexible
fit functions may yield a poor fit radius to better reproduce high-$Q^2$ data.

Another approach that can help address issues of over- or under-fitting data
is the use of the bounded $z$ expansion~\cite{hill06}. In this method, the
form factors are parameterized as a polynomial in $z$ rather than $Q^2$, where
\begin{equation}
z(t,t_{\rm cut},t_0) =
\frac{\sqrt{t_{\rm cut} - t} - \sqrt{t_{\rm cut} - t_0}}
{\sqrt{t_{\rm cut} - t} + \sqrt{t_{\rm cut} - t_0} } \,,
\end{equation}
where $t_{\rm cut}=4 m_\pi^2$ and $t_0$ is a free parameter. The true form
factor is guaranteed to be in the space of this polynomial expansion for large
enough number of parameters, and sum rules exist which limit the size of the
coefficients in the expansion. The details of the $z$ expansion are discussed
for the proton charge and magnetic form factors in Refs.~\cite{hill10,
epstein14, lorenz15, lee15}.

The constraints on the coefficients allow us to estimate bounds which can be
applied to the individual fit coefficients. Applying such a bound to the fit
prevents the uncertainties from growing out of control as more parameters are
added, because it damps the oscillatory behavior that can occur if each term
cancels the error from the previous term. The bounded $z$ expansion
provides fits which are more stable in both the extracted radius and
uncertainty as one increases the number of fit parameters~\cite{hill10,
epstein14, lee15}. This allows the number of parameters to be large enough that
the fit is not limited by the truncation of the Taylor series, while still
providing a meaningful, though perhaps larger, uncertainty on the radii. It
also helps to decouple the parameters needed to fit the low $Q^2$ data from
those important at high $Q^2$, making the fits somewhat more robust against
potential error in measurements at larger $Q^2$ values where the form factor
has little sensitivity to the radius. The estimate of the bounds, however, is
model-dependent~\cite{hill10, epstein14, lee15}, and if the bound applied it
too tight, it can bias the extraction of the radius.

The ability to bound the fit coefficients and go to high order fits is
particularly useful when comparing the analysis of Mainz data and the world
data set, as it allows for the same functional form and number of parameters
to be used in both cases. The charge radii extracted from the Mainz
data~\cite{bernauer13} and world data~\cite{zhan11, sick12} are in good
agreement, but the magnetic radii are significantly different. The fact that
these analyses use very different fit functions and have to select a range of
parameters tailored to the size and precision of the data sets makes it
difficult to determine the role of the model dependence of the fits. With the
$z$ expansion, different data sets, as well as different $Q^2$ ranges of a
single data set, can be examined in a way that minimizes model dependence
associated with choosing different fit functions. A detailed analysis
discussing the remaining model dependence associated with the $z$ expansion
fits and comparing consistent analyses of Mainz and world data has been
undertaken~\cite{lee15}.  This comparison shows that the charge radius from
both Mainz and world data are still inconsistent with muonic hydrogen results,
and that the discrepancy in magnetic radii extracted from the Mainz and world
data persists.

Other analyses have used functions with fewer parameters, requiring only that
the number is sufficient to provide an approximate plateau in the chi-squared
value of the fit and the extracted radii.  This is done because the
uncertainties grow significantly with the number of parameters, and so fits
are generally taken with the minimum number of parameters required to
reasonably fit the data. For the $z$ expansion, the results are
independent of the number of parameters once this becomes large, and so the
most reliable approach is to have several more parameters than is necessary to
obtain a reasonable chi-squared value, to avoid under-fitting of the data. The
fit uncertainties thus tend to be somewhat larger than quoted in previous
results, especially when taking conservative estimates of the coefficient
bounds.  However, the fits should yield more robust estimates of the
uncertainties as they avoid the large dependence of the uncertainty on the
number of parameters used in the fit.

\section{Consistency between Mainz and world data}

As noted earlier, there are significant tensions between the Mainz data set
and world data.  Figure 10 of Ref.~\cite{bernauer13} compares the Mainz
fit to previous world data, and shows a significant disagreement between
the Mainz measurement and nearly all low-$Q^2$ extractions of $G_M$. The 
disagreement in $G_M$ is roughly 3-4\%, corresponding to an 6-8\% cross section
difference if this were explained entirely by normalization factors, as
suggested in~\cite{bernauer13}. However, while the Mainz analysis
yields values of $\gm$ that are several percent above world data, the Mainz
results for $\ge$ are systematically below world data for $0.2 < Q^2 <
0.8$~GeV$^2$. Thus, a simple normalization correction cannot resolve the
discrepancy. Note that the world data results do not have TPE corrections
applied, but at these $Q^2$ values the corrections are relatively small and
for a comparison to the Mainz result, the Feshbach correction should be used.
Applying this correction to the world data would have a small effect that
would decrease $\gm$ and increase $\ge$, further increasing the tension with
the Mainz analysis. A recent analysis~\cite{qattan15} examined both Mainz and
world cross section data, and extracted the TPE contribution using a
phenomenological approach. While the exact TPE extracted at these low $Q^2$
values may not be determined precisely, this procedure applies
a consistent correction to both Mainz and world data and a significant tension
is observed, in particular in the magnetic form factor.

One can also see the disagreement in Figure 19 of~\cite{bernauer13}, which
shows the normalization factors applied to previous measurements as determined
from a global fit. All of the world data sets shown require an increase in 
their normalization, with roughly half of these renormalized by 4\% or more.
This includes several data sets which are shifted by 2-3 times their quoted
normalization uncertainties. As noted earlier, the difference in the way the
uncertainties are separated in the Mainz data will lead to it receiving an
artificially high weight in the fit, but this analysis provides a direct
measure of the large relative renormalization factors required to improve
consistency between Mainz and world data.

\section{Conclusions}

The large number of high-precision data points in the recent Mainz experiment
by the A1 collaboration requires special treatment of the uncertainties. The
release of the data includes uncorrelated uncertainties (statistical and
systematic), correlated systematic uncertainties, and normalization factors
for 34 different subsets of the data. This leads to two concerns with analyzing
these data. First, it is extremely important to account for all of these
uncertainties, in particular the normalization factors and correlated
systematic uncertainties, in any extraction of the radius from these data, as
done in~\cite{bernauer13, lee15, lorenz15}. Second, a simple global analysis
of the Mainz data with other cross section or polarization observable
measurements will give too much weight to the Mainz data, as the uncertainties
given for each cross section point represent only a small fraction of the
total uncertainty.

Given recent measurements supporting the importance of the hadron
structure-dependent TPE corrections~\cite{adikaram15, rachek15}, it is clear
that a $Q^2$-independent TPE correction is not sufficient and that an
uncertainty associated with the model dependence of the TPE correction needs
to be included. We note this and other contributions that are not included in
the evaluation of the correlated systematic uncertainties, but which can have
a significant impact on the uncertainty of the extracted radius. In addition,
different assumptions about the kinematic dependence of these unknown
systematic corrections can noticeably increase the impact of these corrections
on the extracted radii. Between missing contributions to the total systematic
uncertainties and the model dependence of evaluating the impact of these
corrections, it appears that the systematic uncertainties associated with the
radius extraction of~\cite{bernauer13} are likely to be significant
underestimates of the true uncertainty.

Evaluating the model dependence of such fits is important, and while one
can select fit functions and ranges of parameters which appear to yield good
fits to the data with reasonable uncertainties, it is difficult to 
cleanly determine if one is under-fitting or overfitting the data, either
of which can significantly modify the extracted radius and uncertainties.
We argue for the use of the bounded $z$ expansion, which allows a large number
of parameters without concern about overfitting and without the dramatic loss
of precision that comes with unbounded fits at high order. This procedure
yields somewhat larger uncertainties, but yields a more reliable determination
of the uncertainty in the extracted radius.

Initial results from bounded $z$ expansion fits to both the Mainz data
and world data~\cite{lee15} yield consistent charge radii which are still
significantly higher than the muonic hydrogen results. However, there are
clearly inconsistencies between the Mainz data and other world data, both at
the cross section level and in the extraction of the magnetic radius, where
the Mainz data set yields a much smaller magnetic radius.
Without a better understanding of the origin of the tension between the
different data sets, it is difficult to make a clear and rigorous statement
about the present uncertainty on the proton's charge radius as derived from
elastic electron scattering. Based on the considerations presented here and
other examinations of the model dependence of the radius
extractions~\cite{sick12, sick14}, a recommendation for a radius and
uncertainty based on published fits to electron scattering
data~\cite{bernauer13, zhan11, sick12} is presented in
Ref.~\cite{arringtonsick_theseproc}.

This work was supported by the U.S. Department of Energy, Office of Science,
Office of Nuclear Physics, under contract DE-AC-06CH11357

\bibliography{Arrington_Radius}
\end{document}